\documentstyle[epsbox]{elsart}
\begin{document}
\begin{frontmatter}
\title{Origin of Critical Behavior in Ethernet Traffic}

\author{Kensuke Fukuda\thanksref{fff}}
\address{NTT Network Innovation Laboratories, 
3-9-11, Midori-cho, Musashino, Tokyo 180-8585, Japan}
\author{Hideki Takayasu}
\address{Sony Computer Science Laboratories, 
3-14-13, Higashi-Gotanda, Shinagawa-ku, Tokyo 141-0022, Japan}
\author{Misako Takayasu}
\address{Department of Complex Systems, Future University-Hakodate, 
116-2 Kameda-Nakano, Hakodate 041-0803, Japan}

\thanks[fff]{Corresponding author.\\
{\it E-mail address:}\/ fukuda@t.onlab.ntt.co.jp (K.Fukuda)}

\begin{abstract}
We perform a simplified Ethernet traffic simulation in order to 
clarify the physical mechanism of the phase transition 
behavior which has been 
experimentally observed in the flow density fluctuation 
of Internet traffic. 
In one phase traffics from nodes connected with an Ethernet cable 
are mixed, 
and in the other phase, the nodes alternately   
send bursts of packets. 
The competition of  sending  packets 
among nodes and the binary exponential back-off algorithm are revealed to 
play important roles in producing $1/f$ fluctuations at the 
critical point. 
\end{abstract}
\end{frontmatter}

\section{Introduction}
The Internet can be viewed as  an autonomous system in which  
the nodes are heterogeneously connected without any 
central control. 
In the Internet, the unit of information is a packet, 
and many researchers have been investigating the 
statistical properties of packet density fluctuations. 

In 1994 Leland et. al analyzed the time series of  packet 
flow density in the Internet, and showed the existence of 
the $1/f$ type fluctuation\cite{Leland94}. 
Similar to the packet flow fluctuation, 
Csabai  reported that the time series of  round trip time (RTT) 
exhibits the self-similarity in  certain path of the Internet
\cite{Csabai94}.
Following these pioneering works, many observations clearly demonstrate  that 
Internet traffic are characterized by the long-range dependency, 
and the assumption of Poisson process, which had been a major traffic model in 
the traditional traffic theory, has clearly lost its validity
\cite{Paxson95,Crovella97}. 

Takayasu et al. pointed out that 
the Internet traffic can be viewed as a phase transition phenomenon 
between congested and sparse phases by both the RTT experiment
\cite{Takayasu96a,Fukuda99a,Fukuda99b} and 
the packet flow density analysis\cite{Takayasu99a,Takayasu99b,Fukudaphd}. 
Only at the critical point they  found the $1/f$ type fluctuation 
consistent with the above results. 
Namely, they clarified that the self-similarity cannot always be observed in the 
Internet traffic, and the phase transition view is more general and adequate. 

Although there are many observational evidences for this phase transition, 
we still do not fully understand the physical explanation of the observed  
phenomena. 
Takayasu et al. reported that a simple queue itself plays an important role 
in phase transition phenomena in general
\cite{Takayasu97}. 
We believe that the theory is qualitatively correct, however, 
there are some quantitative differences between the observation and the 
simple queueing theory. 
For example, 
the congestion duration time is known to be characterized by 
the power law distribution whose exponent is $-1.0$ at the critical point, 
while the simple queue model 
can only reproduce the traffic characterized by the power law distribution 
whose exponent is $-0.5$  at the critical point. 
Thus, we need to find a mechanism that causes such 
difference in exponents.

In this paper, 
we focus on the 
effect of the Ethernet (CSMA/CD) mechanism 
in order to give 
a  more sophisticated physical explanation 
for the observation facts. 
Ethernet has been mostly 
used in the local area network in the Internet. 
Ethernet itself,  however, has  very complicated mechanisms to  
achieve  efficient communication, so 
it is difficult to clarify the role of each mechanism  
in the phase transition. 
For this reason, we perform a simulation based on the  minimal mechanism of 
the Ethernet algorithm, especially focusing on the two effects; 
the competition among nodes that intend to send their packets to 
the shared media, and 
the exponential back-off algorithm in collision detection. 

We review the observation facts in the following section. 
In section 3 we introduce a model of Ethernet traffic with 
simplified algorithm for numerical simulation.
Simulation results are described in Section 4 
which is divided into four sub-sections.
We discuss the cause of 1/f type fluctuations in section 5. 
The final section is devoted to the summary.

\section{Observation of Phase Transition Phenomena in Ethernet Traffic}
\subsection{Data Measurement Environment}
In order to collect the raw data trace, 
we set up a FreeBSD PC connected to the 10Mbps Ethernet link 
between the WIDE (Widely Distributed Environment project) backbone and 
Keio University in Japan by a non-intelligent hub. 
There are no other hosts in this link,  
so the tapping host can capture all packets and the time stamps through this 
link by using the tcpdump command. 
From the measurements, we obtained  12 data traces, each 14440 seconds 
(about 4 hours) long  between Nov. 1997 and Feb. 1998.  
Each trace is categorized into the 
three typical time periods, i.e. early morning, business hour, and 
evening. 
We reconstruct each original tcpdump trace 
into seven time sequences of the flow density fluctuation in bytes, 
whose sample size is $0.1$ seconds.  

\subsection{Phase Transition Phenomena in Real Traffic}
Figure 1 shows an example of packet flow density fluctuation of a 4-hour 
measurement. 
\begin{figure}
\begin{center}
\epsfile{file=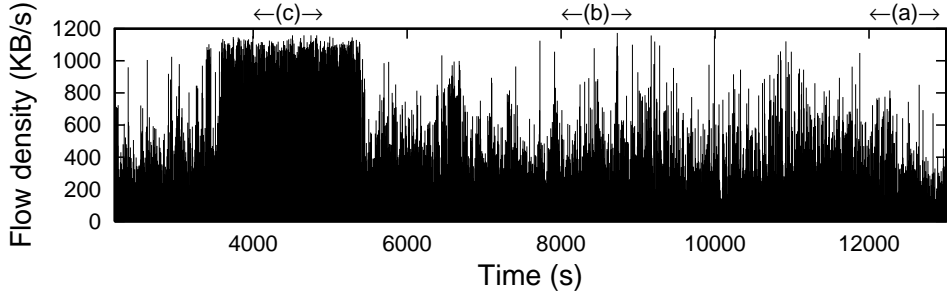,scale=0.6}
\caption{Real Ethernet information traffic fluctuation}
\end{center}
\end{figure}
The abscissa shows the time in seconds and the ordinate is 
the corresponding flow density in bytes. 
The periods denoted by (a), (b), and (c) in the figure indicates  
three typical periods, namely, sparse, moderately-congested, and 
congested periods, respectively. 
The mean flow densities of the three periods are 150 kbyte/sec, 550 kbyte/sec, 
and 850 kbyte/sec, respectively. 
From the figure, it is clear that  the flow density highly fluctuates in the 
moderately-congested periods (b) in Figure 1, 
though it roughly stays in lower or higher states in the other cases, 
(a) and (c), respectively.

Next we explain the data analysis  method called as 
congestion duration length analysis\cite{Takayasu93b}. 
Congestion duration length is a major index to 
characterize the statistical nature of a given time fluctuation. 
We first define a congestion state that 
the flow density has larger value than a threshold value. 
Then, the congestion duration length is defined as 
the sequential number of the congestion state multiplied by the bin size of 
the original time series. 
We are interested in the cumulative distribution of this congestion duration 
length. 
At the critical point, it is known that 
the congestion duration length distribution follows the 
power-law distribution with exponent $-1.0$. 
This slope is theoretically corresponding to the so-called $1/f$ power spectrum in the 
original time series. 

Figure 2 shows the congestion duration length distribution 
in a log-log plot. 
\begin{figure}
\begin{center}
\epsfile{file=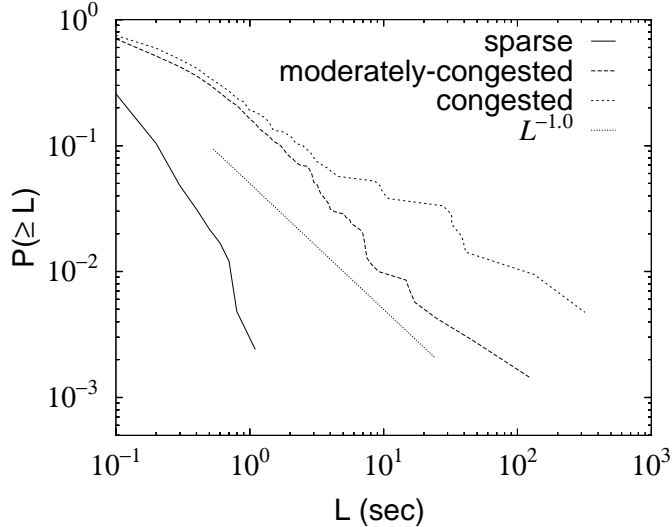,scale=.8}
\caption{Congestion duration length distribution}
\end{center}
\end{figure}
The flow data traces of these three plots correspond to the three periods 
shown in (a), (b), and (c) in Figure 1. 
The plotted line clearly decays exponentially  
in the sparse period, 
and the distribution is approximated by a power law distribution with exponent 
$-1.0$ in the moderately-congested period. (The straight line in this figure 
represents the power law distribution with slope  $-1.0$.)
Thus, this observed power law distribution shows that the mean flow 
density in this period is close to the critical point.
Moreover, above the critical point, the distribution deviates from the 
power law  again due to large  clusters of congestion. 

From these figures we confirm the existence of the typical two phases, 
namely, the sparse and  congested phases. 
Particularly in the intervals, which are close to this critical mean 
flow density,  the congestion duration 
length distribution clearly shows the self-similarity. 
Also, it is reported that the tendency of divergence of the autocorrelation 
time the whole behaviors can be confirmed near this critical mean flow density in the same data 
traces\cite{Takayasu99a,Fukudaphd}. 
Namely, the actual Ethernet traffic flow generally changes its statistical 
property, which can be fully modeled by the phase transition view.

\section{Simplified Ethernet Traffic Simulator}
In this section, 
we focus on the dynamic aspect of the Ethernet mechanism, 
in order to clarify the physical mechanism of the observed 
phase transition phenomenon.
Since the actual Ethernet dynamics consists of many complicated rules,  
it is difficult to extract the essence of the phase transition 
phenomenon directly from  a real system.
Therefore, we need to focus on a few basic properties of Ethernet rules by 
introducing a minimal network architecture model to check the occurrence of the 
phase transition. 
For this purpose, we consider the two most basic effects 
in the Ethernet mechanism; the competition of the nodes at  packet 
transmission,   and the binary exponential back-off algorithm. 
Also, to simplify the simulation, we assume  
a topology  consisting of two nodes (to be called as nodes 1 and 2) 
sending  packets randomly and one shared medium to connect them.

Figure 3 illustrates our simplified discrete simulation algorithm. 
\begin{figure}
\begin{center}
\epsfile{file=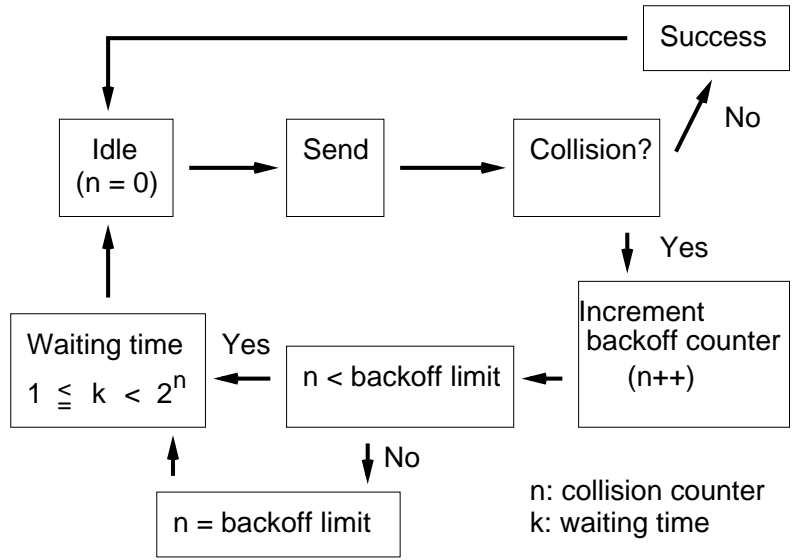,scale=0.8}
\caption{Simplified Ethernet simulation algorithm}
\end{center}
\end{figure}
In every discrete time step, each  node probabilistically 
generates a fixed-size packet to be sent to the medium. 
Each node has an assigned input rate, and 
a packet is created randomly with this rate and injected into the finite-size 
output queue of this node. 
This input rate is given initially and it does not change during 
the simulation. 
As known directly from this rule of  packet creation, 
there is no temporal correlation among the input traffic to the queue. 

The maximum size of the output queue is 16384 ($= 2^{14}$) packets 
in this simulation. 
When a generated packet finds no room in the queue, 
it is simply discarded.
If more than one  packets are waiting in the queue, 
then the node tries to transmit a packet from the queue to the shared network 
medium.
When only one node intends to send a packet 
to this network, this node successfully transmits one packet from the queue, 
and it takes one time step to finish its transmission. 
However, when more than one node try to send  packets simultaneously,  
both nodes fail to transmit them,  and the packets  remain in 
their queues. 
Then, the nodes increase their back-off counters, which indicate the level of 
the back-off, by 1.
Furthermore, they set their own waiting time $k$  
chosen randomly from [1, $2^{n}$] where $n$ denotes the back-off counter. 
Each waiting time decreases one time step in every time step, and 
a waiting node tries to send a packet when 
its waiting time becomes zero. 
The successful node, conversely, resets its back-off counter to zero after 
transmission. 

In our simulation the maximum level of the back-off counter is set 
to  $14$,  i.e. the waiting time is chosen from a random number smaller 
than $16384$ ($= 2^{14}$). 
Also, the back-off counter remains the maximum number even after  
the node fails  to send a packet more than $14$ times successively, while 
the actual Ethernet algorithm drops the packet when the back-off limit 
is larger than the maximum back-off number. 
It should be emphasized that new packets are stored in the queue 
corresponding to  the input rate even when the node is in the waiting status.

\section{Observation of Phase Transition Phenomenon}
Here, 
we focus on the property of the total amount of packet 
traffic flow passing through the link based on our simplified Ethernet 
algorithm.
\subsection{Two-node Case}
In Figure 4 we plot macroscopic performance metrics
(packet dropping rate, throughput, and reliability) in our simulation. 
The abscissa denotes the input rate of the nodes, where we set the 
same input rate for both nodes. From the definition of input rate, 
the input rate 50\% statistically corresponds to the 
maximum link capacity. 
\begin{figure}
\begin{center}
\epsfile{file=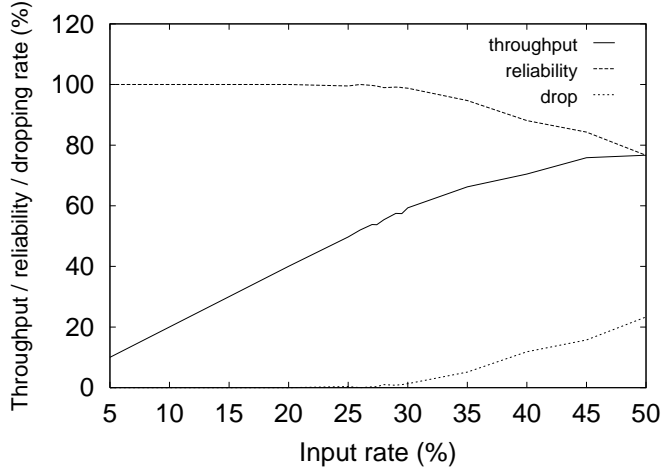,scale=0.7}
\caption{Performance metrics (packet dropping / throughput / reliability) 
vs. Input rate}
\end{center}
\end{figure}
Here, the packet dropping rate is given by the ratio of the number of 
packets overflowing at either of the nodes divided by the total number of 
input packets. The throughput is defined as the number of packets 
normalized by the link capacity. The reliability shows 
the rate of successfully transmitted packets, which is given by 
100\% minus the packet dropping rate. 
Obviously, packet dropping rate suddenly takes non-zero values for input rate 
above about 30\%, 
which is to be called as the critical input rate. 

Figures 5 (a), (b), and (c) show three typical packet flow fluctuations 
for input rates 15, 29.5, and 45 \%, respectively. 
\begin{figure}
\begin{center}
\noindent\epsfile{file=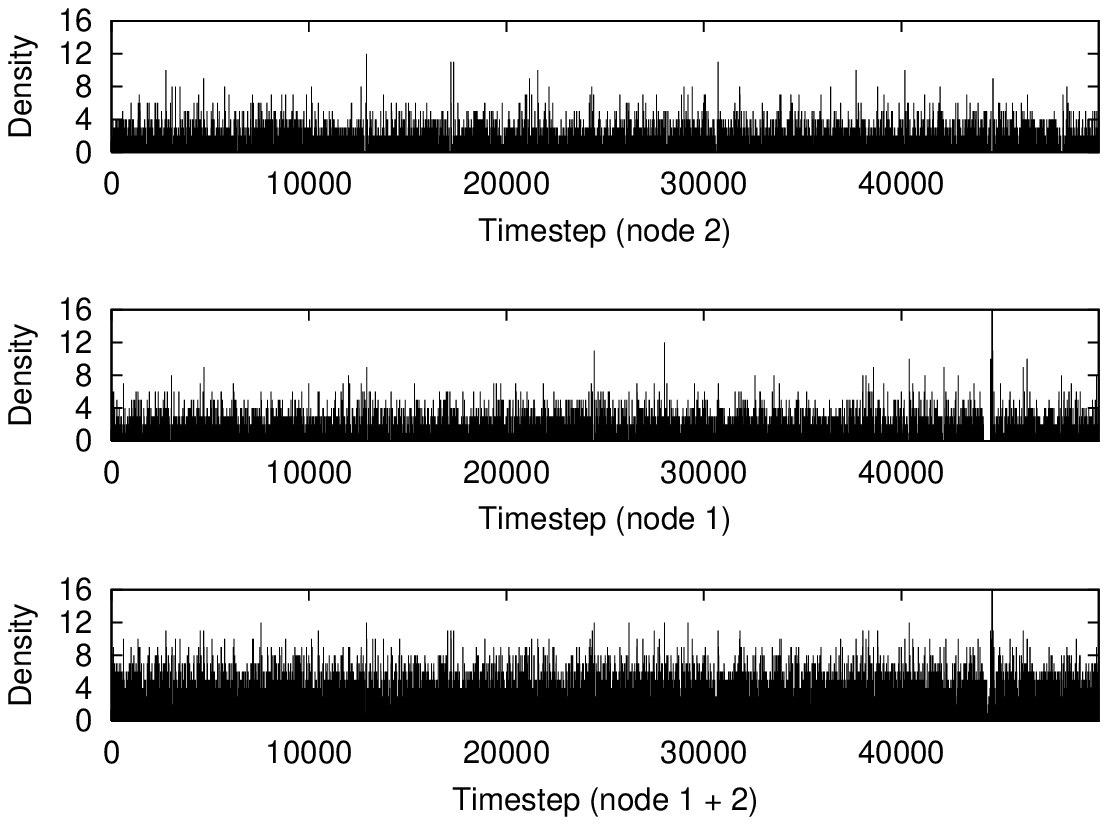,scale=0.7}

\noindent\epsfile{file=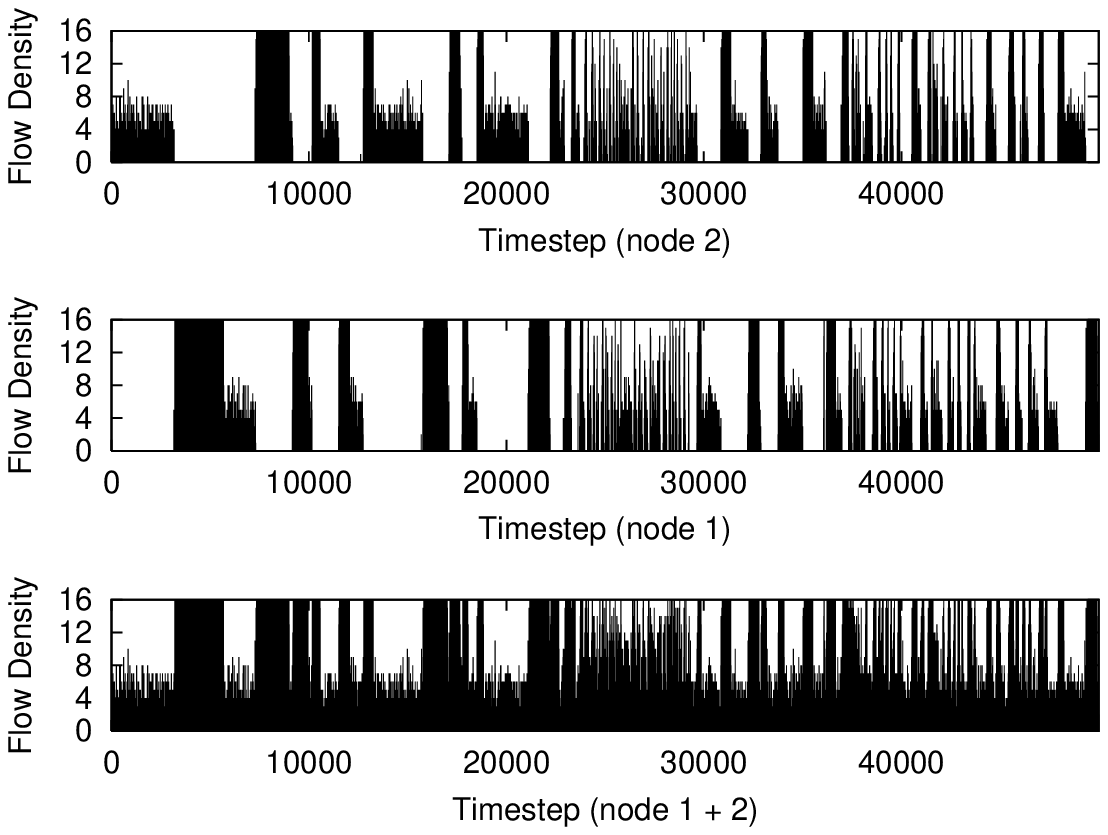,scale=0.7}

\noindent\epsfile{file=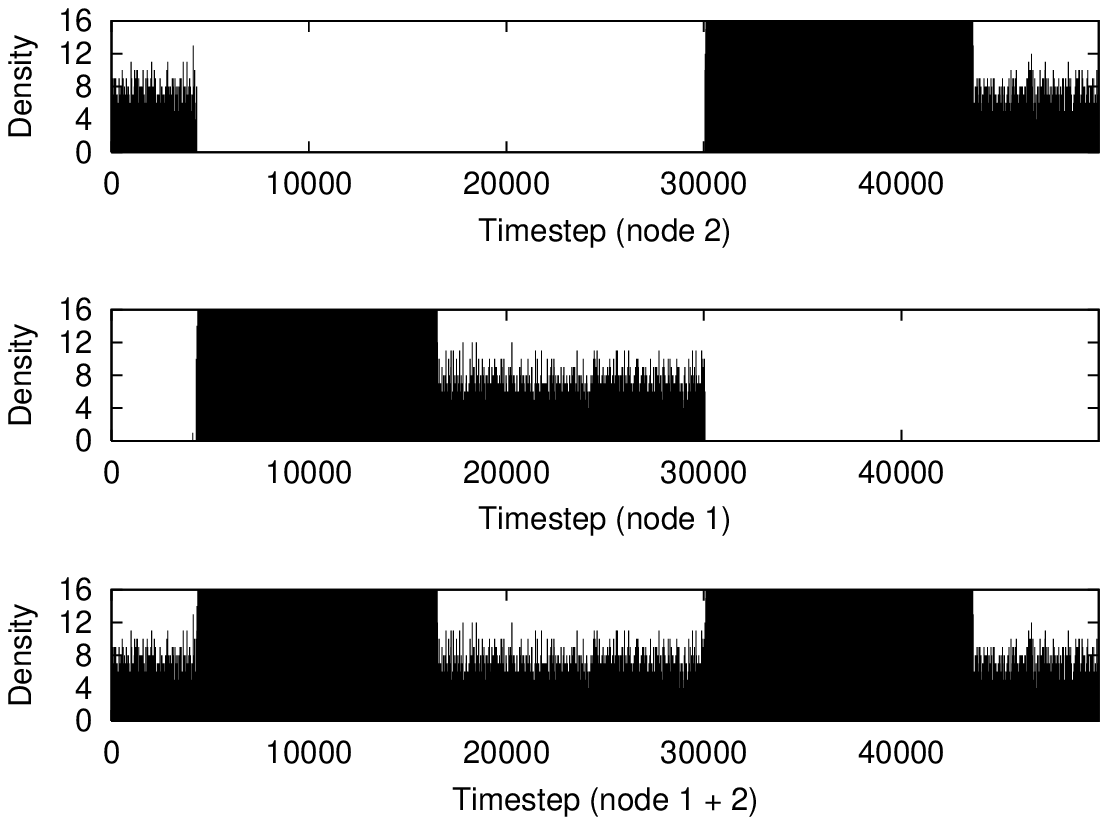,scale=0.7}
\caption{Simulated traffic flow. top (a): input rate = 15\%, 
middle (b): input rate = 29.5\%, bottom (c): input rate = 45\%. 
The sub figures represent the traffic flow from node 1, from node 2, 
and the total flow through the link.}
\end{center}
\end{figure}
Each figure consists of three sub figures, namely,  the output flow 
fluctuations from node 1, from 2, 
and from 1 $+$ 2 (the total traffic in this link). 
The abscissa shows the time step, and the ordinate indicates the 
number of output packets per every 16 time steps. 

At the  low input rate 
the two nodes randomly transmit their packets, and 
the total traffic seems to have  spontaneous small bursts. 
Near the critical input rate,  
the total packet fluctuation consists of  two typical types of 
fluctuations. One is a highly variable fluctuation observed 
around 28000 time steps, where 
the two nodes mutually send packets to the network with 
fine granularity, 
and the total traffic is nearly a superposition of such individual 
traffic flows. 
The other type is observed around 5000 time steps in which one node 
continuously sends  packets until its buffer becomes empty. 
Thus, we can expect the sizes of the congestion to be distributed 
widely at the critical mean flow density. 
In the case that the input rate is high, 
each node alternately sends bursts of roughly same size depending on the 
buffer size.  
The short fluctuations followed by the full bursts observed around 
20000 time steps in node 1 are likely due to 
the imbalance of the buffer size and the maximum back-off waiting time. 
Comparing  the figures, we found that 
the simulated packet flow captures the qualitative characteristics of the 
actual traffic flow. 

Next, we show the result of the congestion duration length distribution in 
this two-nodes simulation in Figure 6. 
\begin{figure}
\begin{center}
\epsfile{file=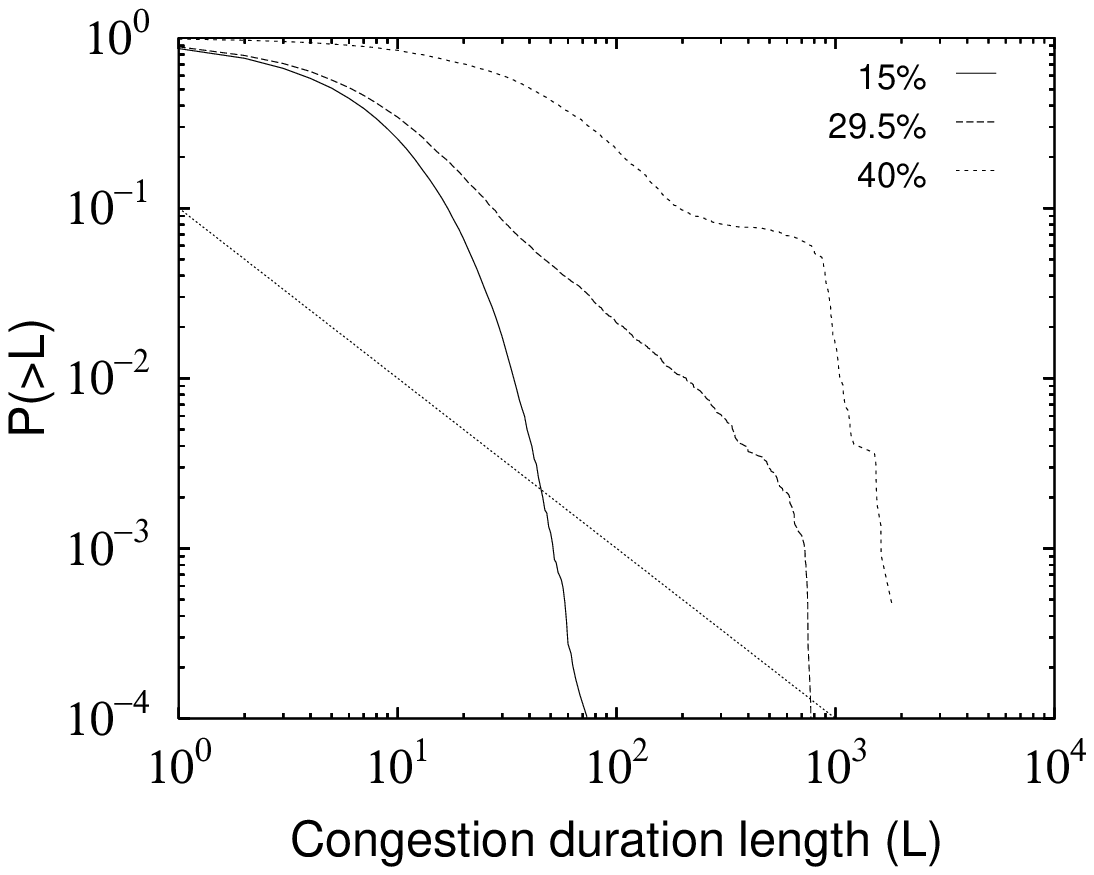,scale=0.8}
\caption{Congestion duration length distribution (2 node). 
The straight line indicates the power law with exponent $-1.0$}
\end{center}
\end{figure}
The original data trace is the total flow density fluctuations 
from the two nodes per 16 time steps, 
and the threshold value of congestion level is set to 
be two packets. 
The three lines in this figure represent the low  (15\%), the 
medium (29.5\%), and the high input rates (40\%) in 
log-log scale. 
When the packet input rate is low, 
the congestion duration length follows the exponential distribution. 
Namely, the congestion duration has only short-range dependency. 
In the case of high input rate 
the distribution  deviates from the power-law distribution 
due to the large clustered congestion. 
The decay of this plotted line at large $L$ is  due to system size 
 limitation (e.g., buffer size, simulation time, etc.). 
In the medium input rate case 
the plotted line is approximated by a power law distribution with exponent 
$-1.0$. 
This input rate is the same as the critical value at which a 
node begins dropping  packets due to the buffer overflow in Figure 4.

From these results,  we  
confirm that the simplified packet traffic model can reproduce the two 
typical phases and the critical point characteristics 
observed in real systems. 
The three types of distributions of the congestion duration lengths are 
consistent with the distributions obtained from the actual 
traffic measurement\cite{Takayasu99a,Fukudaphd}. 
It should be noted that the packet input to the queue 
is based on purely random events of Poisson process, 
so, 
the self-similarity observed in the output fluctuation at the critical point is 
caused by the evolution dynamics 
based on the competition and the binary exponential back-off algorithm.

\subsection{Three-node Case}
The observed network introduced in Section 2 simply consists of only 
two nodes. 
In an ordinary network, 
however, there are generally  many nodes connected to the shared media.  
Therefore, we need to consider the situation in which 
more than two nodes participate.
Here, we show the traffic fluctuations passing 
through a link connected to three nodes.
Figure 7 represents the congestion duration length distributions  
in the aggregated traffic from the three nodes 
following the same simulation algorithm as in the preceding sub-section. 
\begin{figure}
\begin{center}
\epsfile{file=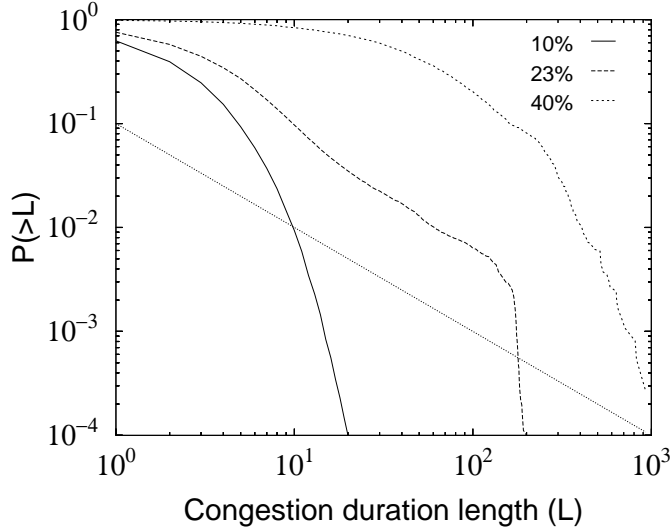,scale=0.8}
\caption{Congestion duration length distribution (3-node). 
The straight line has  slope $-1.0$.}
\end{center}
\end{figure}
Though the critical input rate (23\%) is lower than the case of the two-node 
simulation (29.5\%), the  same three typical types of distributions 
are clearly observed also in this figure. 
Similarly, we found that the three performance metrics (packet dropping rate, 
reliability, and throughput) bend at this critical point. 
We also observed the same type of the phase transition in the four node 
simulation. 
The cases with more than four nodes  give the same result 
in general. 
These results clearly show that the occurrence of the 
phase transition behavior is considered to be independent of the number of nodes connected to the link.

\subsection{Asymmetric Case}
In the previous simulations, we have assumed that 
all nodes had the same input rate. 
Here, we analyze an asymmetric case  
consisting of two nodes having different input rates. 

Figure 8 depicts the phase diagram of the estimated critical point when 
the combination of the input rates of the two nodes changes. 
\begin{figure}
\begin{center}
\epsfile{file=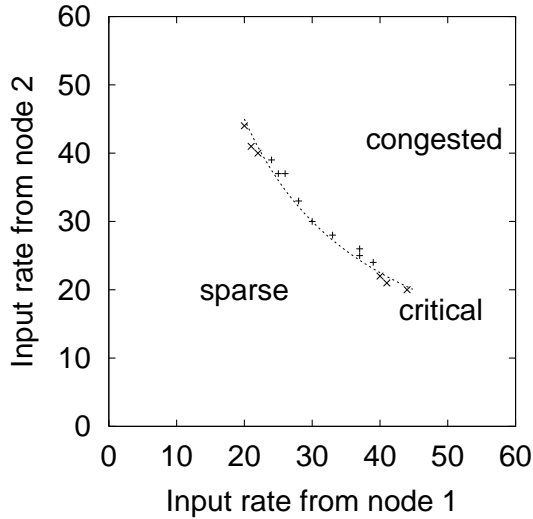,scale=0.8}
\caption{Phase diagram of two asymmetric inputs}
\end{center}
\end{figure}
The abscissa and the ordinate represent the input rates of nodes 1 and 
2, respectively. 
The point (30\%, 30\%) corresponds to the critical point 
in the previous symmetric case. 
The plotted line represents the set of critical points estimated by the 
simulation,  where self-similar traffic flows can be observed.  
We can divide this diagram into two regions 
separated by the critical line.
The aggregated traffic belongs to the non-congested phase 
if the combination of input rates of the two nodes is in the 
lower-left region of this figure, 
and to the congested phase if it is in the upper-right region above 
the critical line.
Their aggregated traffic fluctuations do not show  
self-similarity, and their congestion duration length distributions  
follow approximately an exponential distribution and  
a distribution with a plateau, respectively, as we saw in Figure 6. 

It is found that the phase transition behavior disappears 
when the nodes connected to the link have extremely different input 
rates. 
This can be understood as  follows.
In the extremely asymmetric case, 
the probability of collision, which is proportional to the multiplication 
of the input rates of the two nodes, is 
relatively small. 
As a result, the back-off becomes less effective and 
the statistical property of the input process is likely to be preserved. 

\subsection{Dependence on Back-off Functions}
In order to clarify the effect of 
the functional form of the back-off function, 
we performed the simulation with a different  back-off rule. 
With the normal Ethernet algorithm, when collision is detected, 
the waiting time is chosen randomly in the interval between 1 and $2^{n}$,  
where $n$ is the back-off counter. 

For comparison with this basic algorithm, 
we check the case of a linear back-off algorithm in this simulation, 
namely,  
the back-off time 
is  selected from a random time between $1$ and $n * k$, where 
the coefficient $k$ is set to be $100$ in this simulation. 
We have confirmed that the following results are the same for $k$ 
between $10$ and $1000$. 

Figure 9 shows the results of the congestion duration length analysis 
for the linear back-off algorithm.  
\begin{figure}
\begin{center}
\epsfile{file=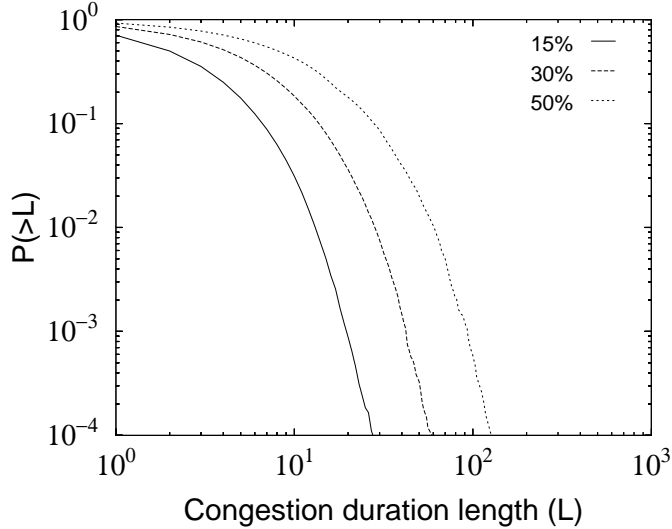,scale=0.8}
\caption{Congestion duration length distribution with linear back-off}
\end{center}
\end{figure}
The plotted lines clearly follow  exponential functions independent of 
the value of the input rate. 
Thus, neither the phase transition phenomenon nor the self-similarity 
can be observed in the resulting traffic. 
It is now clear that the binary exponential back-off algorithm plays an important 
role in generating self-similar traffics. 

Similarly, we also check the case of 
ternary exponential back-off algorithm, 
that is, the waiting time $k$ is chosen randomly from [$1$, $3^n$] with 
$n$ being the back-off counter. 
By this simulation 
we find the power law distribution with the same slope $-1.0$ at the 
critical point 
while the critical point itself shifts from the value in the 
binary exponential back-off case. 
Therefore, the exponential type back-off function is expected to be 
the essence of the observed phase transition behaviors. 

\section{Discussion}
We have found that our simplified Ethernet simulation 
based on  both competition in transmission 
and the exponential back-off algorithm 
can reproduce essentially the same traffic behaviors as observed in  
real Ethernet traffic,  
When the total traffic flow rate is low, 
there are few collisions in the link, 
and the  traffic statistics are dominated by the random 
input. The aggregated traffic is approximately consisted of 
superposition of the input traffic from the two nodes. 
However, at the critical input rate, 
the  number of collisions becomes not negligible and the back-off mechanism works 
effectively, and the resulting output traffic becomes correlated 
in long time scales.
The traffic is hovering about the 
two phases rather randomly causing large fluctuations showing  
self-similarity.
Above the critical input rate  
the traffic is dominated by  
clustered congestion whose size depends both on the buffer size of the nodes and 
on the maximum number of the back-off count. 
In this phase the traffic loses the self-similarity. 

In the single queue model the power law exponent for 
the congestion duration length distribution 
is known to be $-0.5$ at the critical point\cite{Takayasu96a}. 
Our results show that the network traffic behavior from more than two nodes 
following  the simplified back-off algorithm exhibits a  power law distribution 
with exponent $-1.0$. 
Thus, we conclude that the back-off delay due to competition among the nodes is a key factor 
in generating the $1/f$ type fluctuation.
In particular, it is very important that 
the phase transition and resulting self-similarity appear 
in the output traffic behavior even when the input traffic 
is purely random, having no temporal correlation. 

Moreover, our simulation demonstrates that the linearly incremental back-off 
algorithm fails in realizing a  
phase transition. Thus, it is clear that the 
multiplicative form of the back-off function 
is an essential factor in characterizing the statistical property of the 
Ethernet traffic behavior. 

Finally, our simulation demonstrates that 
the critical point does not always exist when the 
nodes have extremely  different input rates. 
In the highly asymmetric case 
packets have less chance to collide with other packets, 
and the  back-off value cannot become very large. 
For this reason the output traffic is not much modified from 
superposition of input traffics. 
We can expect that 
the critical behavior can be found in a wider region in the phase diagram 
if  the input traffic is temporally correlated because 
correlated packets generally have more chance of collision with other packets 
than the case of independent random inputs. 

\section{Conclusion}
Our simulation is based on only two effects in the complicated 
Ethernet mechanism; the competition among the nodes 
at transmission process and the  
exponential back-off. 
The simplicity of the simulation algorithm enables us to investigate detailed 
statistical properties of packet traffics. 

Our results can be summarized as follows: 
(1) The simplified simulation can reproduce the basic properties of the 
real traffic fluctuations especially in terms of the phase transition 
between the sparse or mixed transmission phase and congested alternating transmission 
phase. 
(2) At the critical flow density the output flow shows a self-similarity in 
its fluctuation even though the input is totally random without having 
any temporal correlation. 
(3) Phase transition does not appear when the input rates of the nodes are 
extremely asymmetric. 
The approximate phase diagram in the parameter space is drawn with a 
critical curve. 
(4) The occurrence of the phase transition is expected to be independent of the 
number of the nodes. 
(5) The functional form of the back-off algorithm plays an essential role. 
The linear back-off function can not realize a phase transition, 
while the exponential back-off always reproduce the 
$1/f$ type fluctuation at the critical point. 

\section{Acknowledgments}
We wish to thank K. Cho and Y. Takeshima for helpful discussion.

\end{document}